\documentstyle[aps,epsf,eqsecnum]{revtex}

\newcommand{\beq}{\begin{equation}}
\newcommand{\beqn}{\begin{eqnarray}}
\newcommand{\eeq}{\end{equation}}
\newcommand{\eeqn}{\end{eqnarray}}

\newcommand{\ts}{  \textstyle}

\newcommand{\gsim}{\mbox{\raisebox{-1.ex}{$\stackrel
     {\textstyle>}{\textstyle\sim}$}}}
\newcommand{\lsim}{\mbox{\raisebox{-1.ex}{$\stackrel
     {\textstyle<}{\textstyle \sim}$}}}
\newcommand{\square}{\kern1pt\vbox{\hrule height
1.2pt\hbox{\vrule width 1.2pt\hskip 3pt
   \vbox{\vskip 6pt}\hskip 3pt\vrule width 0.6pt}\hrule
height 0.6pt}\kern1pt}
\def\({\left(}
\def\){\right)}
\def\[{\left[}
\def\]{\right]}

\def\tsig{\tilde{\sigma}}

\def\cb{{\cal B}_k}

\begin{document}
\draft
\twocolumn[\hsize\textwidth\columnwidth\hsize\csname
@twocolumnfalse\endcsname

\title{\bf Preheating - cosmic magnetic dynamo ?}

\author{ 
Bruce A. Bassett,${}^{1}$ Giuseppe Pollifrone,${}^{2,3}$
Shinji Tsujikawa${}^{4}$ and Fermin Viniegra${}^{5}$}
 \address{${}^{1}$ 
Relativity and Cosmology Group, School of Computer Science and Mathematics, 
Portsmouth University, Portsmouth~PO1~2EG, England} 
\address{${}^{2}$ 
Astronomy Unit, School of Mathematical Sciences, Queen Mary,
University of London, London~E1~4NS, England} 
\address{${}^{3}$ 
 ABN AMRO Bank N.V.,  250 Bishopsgate
London, EC2M 4AA}
\address{${}^{4}$ Department of Physics, 
Waseda University, 3-4-1 Ohkubo, Shinjuku-ku, Tokyo 169-8555, Japan} 
\address{${}^{5}$ Department of Theoretical Physics, Oxford University, 
Oxford~OX1~3NP, England}
\date{\today}
\maketitle
\begin{abstract}
We study the amplification of large-scale magnetic fields during
preheating and inflation in several different models. Preheating can 
resonantly amplify  seed fields on cosmological scales. 
In the presence of conductivity, however,  the effect of resonance 
is typically weakened and the amplitude of produced magnetic fields 
depends sensitively on the evolution of conductivity 
during the preheating and thermalisation phases.
In addition we discuss geometric magnetisation,
where amplification of magnetic fields occurs through coupling to
curvature invariants. This can be efficient during inflation  
due to a negative coupling instability. 
Finally we discuss the breaking of the conformal flatness of the 
background metric whereby magnetic fields can be stimulated through the 
growth of scalar metric perturbations during metric preheating.
\end{abstract}
\pacs{98.80.Cq \hspace*{0.2cm}}
\centerline{PU-RCG-00/33, WUAP-00/27, astro-ph/0010628}
\vskip 1pc
 ]

\section{Introduction}

With the current dominance of the inflationary paradigm and  the
gravitational instability picture of structure formation seeded by quantum
fluctuations it is easy to forget earlier, competing,  models. In
particular, models of structure formation based on turbulence had the
advantage that they were able to make strong connections between galaxy
formation, galactic angular momentum and galactic magnetic fields
\cite{wass78}.

Inflation, by contrast, predicts essentially zero vorticity and
in its purest forms,\footnote{With no explicit terms or interactions
which break conformal invariance.} rather small magnetic fields.
The end of inflation may be very violent, with rapid particle
production -- a process known as preheating. During preheating, 
fluctuations of scalar and Gauge fields exhibit exponential growth by parametric resonance \cite{earlypre,KLS,on}. It has a host of potentially radical 
side-effects: 
Grand Unified Scale baryogenesis \cite{baryogenesis}, non-thermal symmetry 
restoration \cite{ntsr}, and topological defect formation \cite{defect}.  
Here we will discuss a side effect which may have persisted until the 
present day - the amplification and sculpting of primordial magnetic fields 
to the amplitudes seen today on cosmic scales.

Magnetic fields are known, partly via the Faraday rotation of light they
induce, to permeate many astro-physical systems including intra-cluster
gas, quasars, pulsars and spiral galaxies. The fields are large, with
magnitudes $\sim 3 \times 10^{-6}$ G on scales greater than $10$ kpc
\cite{review}. Such amplitudes present an ``inverse" fine-tuning problem 
as compared with the standard one 
in inflation\footnote{For example, for the
potential $V = \frac14 \lambda\phi^4$, CMB anisotropies in the absence of
preheating demand $\lambda \sim 10^{-13}$, a rather severe fine-tuning.}:
Since Maxwell's equations are conformally invariant and 
Friedmann-Lema$\hat{\i}$tre-Robertson-Walker (FLRW) models are conformally
flat\footnote{i.e., The Weyl tensor, $C_{\alpha\beta\mu\nu}$, vanishes.}, the 
cosmic expansion does {\em not} create photons or 
magnetic fields.  The origin of 
these large amplitude fields, correlated on such large scales, is still 
generally regarded as an unsolved mystery, despite the proliferation of 
putative explanations \cite{parker,ZRS}.

The observed magnetic fields today have an energy density comparable to
that in the Cosmic Microwave Background (CMB): $r \equiv B^2/(8\pi
\rho_{\gamma}) \sim 1$ \cite{TW}. If we run the cosmic clock backwards
past a redshift of $z > 100$ where structure formation is strongly in the
linear regime, $r$ may have decreased to around $10^{-34}$ through the
combined effects of the galactic dynamo \cite{parker,ZRS} and collapse of
structure, which amplifies the magnetic field as $(\delta\rho/\rho)^{2/3}$
due to flux conservation. The galactic dynamo efficiently converts
differential rotation of spiral galaxies into magnetic field energy and
without it $r \sim 10^{-8}$ is required to seed the observed fields \cite{TW}.

The limit on a homogeneous magnetic field on horizon scales today is 
$~\lsim~10^{-9}$ G \cite{BFS}. In contrast, at decoupling a magnetic field at 
smaller scales would lead to dissipation of energy into the photon 
fluid and  lead to spectral distortions.  To avoid  conflict 
with COBE FIRAS results requires the field to be  less 
than $\sim 10^{-8}$ G today at scales $0.4 - 600$ kpc.

The time evolution of $r$ is typically believed to be
rather trivial: $r \sim$ constant. This is due to the high conductivity of
the universe through the matter and radiation dominated phases which
conserves magnetic flux and leads to the behaviour $B \sim a^{-2}$ and
$B^2/\rho_{\gamma} \sim $ constant. However, during preheating and
inflation, the low conductivity of the universe, due to the paucity of
charged particles, creates an environment in which $r$ can change freely.

The production of magnetic fields during inflation has been studied by
Turner and Widrow \cite{TW} and Davis {\em et al.} \cite{DDPT} and during 
phase transitions by several authors \cite{Vachaspati,EO,DD}.  In reheating 
their production via stochastic currents was investigated by Calzetta {\em et 
al.} \cite{calzetta}.  

In this paper we consider the mechanisms discussed by Turner and Widrow
\cite{TW} and show  how preheating may lead to resonant amplification
of magnetic fields \cite{FG}.
We also discuss a mechanism \cite{BGMK,maroto} based on the breaking of 
conformal flatness of the background geometry due to metric preheating 
rather than breaking of the conformal invariance of the Maxwell equations.  
Although they also lead to resonance, we do not consider the axion-like 
couplings $\phi F_{\mu\nu}^*F^{\mu\nu}$ since they have been considered in 
depth by a number of authors \cite{Car,Bru}.  We will also not describe 
resonant production of magnetic fields in low-energy string actions where 
conformal invariance is broken by the existence of the dilaton $\phi$.  
Such models have been discussed in 
\cite{ratra,dolgovanomaly,lemoine,gasperini}.

\section{Magnetic fields in curved spacetime}

Maxwell's equations arise from the Lagrangian density $-\ts{1\over 4}
F^{\mu\nu}F_{\mu\nu}$, where $F_{\mu\nu} \equiv 2 
\nabla_{[\mu} A_{\nu]}$ is the Maxwell tensor, 
$A_{\mu}$ is the four-potential, $\nabla_{\mu}$ is
the curved space, covariant derivative, and square brackets on
indices denote anti-symmetrisation on those indices.

The Maxwell equations that arise are then:
\beq
\square A_{\mu} + R_{\mu\nu} A^{\nu}-
\nabla_{\mu} \nabla_{\nu}A^{\nu} = 0\,,
\label{maxeq}
\eeq
where $\square \equiv \nabla_{\mu} \nabla^{\mu} =
(1/\sqrt{-g})\partial_{\mu}(g^{\mu\nu}\sqrt{-g}\partial_{\nu})$ and 
$g \equiv {\rm det}(g_{\mu\nu})$. The Ricci tensor term arises through the
non-commutativity of covariant derivatives and application of the
contracted Ricci identities $2\nabla_{[\mu\nu]}A^{\nu} =
R_{\mu\nu}A^{\nu}$ \cite{ellis}.

The four-potential suffers from a gauge freedom which must be eliminated.
One may use either the covariant Lorentz gauge condition $\nabla^{\mu} A_{\mu}
= 0$ or the combined Coloumb/tri-dimensional/radiation  gauge  conditions
$A_0 = 0, \partial^i A_i = 0$.  In both cases the last term in Eq.  
(\ref{maxeq}) vanishes \footnote{If one explicitly breaks the $U(1)_{\rm EM}$ 
gauge invariance and conformal invariance by introducing a photon mass term 
$m^2 A_{\nu} A^{\nu}$ into the Lagrangian, then one recovers the Proca 
equation, and the gauge condition $\nabla^{\mu} A_{\mu} = 0$, becomes a 
true constraint equation.}.

Except for the last section we will use a flat FLRW spacetime 
as a background. 
The metric is then
\beq
ds^2 =  a^2(\eta)(-d\eta^2 + \delta_{ij}dx^i dx^j)\,,
\label{FLRW}
\eeq
where $\eta \equiv \int a^{-1}dt$ is conformal time, $a(\eta)$ is the scale
factor of the universe and $\delta_{ij}$ is the Kronecker delta.
The traceless part of the Riemann tensor -- the Weyl tensor
$C_{\alpha\beta\mu\nu}$ -- defined by \cite{ellis}, 
\beq
C_{\alpha\beta\mu\nu} = R_{\alpha\beta\mu\nu}
-g_{\alpha[\mu} R_{\nu]\beta}+g_{\beta[\mu} R_{\nu]\alpha}
+\frac13 R g_{\alpha[\mu} g_{\nu]\beta}\,,
\label{weyl}
\eeq
vanishes in FLRW backgrounds which are therefore conformally flat. The
metric (\ref{FLRW}) is also conformally static.

Placing a {\em homogeneous} magnetic field in a FLRW background is not
consistent since the magnetic field picks out a preferred direction which
is not consistent with the maximal symmetry spatial subsections of the
FLRW models. Instead, the (anisotropic) Bianchi models provide an
appropriate background for the study of this problem
\cite{TM}.

Instead we will assume that the magnetic field produced will not be
coherent on very large scales. Such a possibility is already strongly
constrained by the CMB. Rather we will assume that the  power
spectrum, $B(k)$, of the magnetic field is statistically isotropic and
homogeneous, hence consistent with the symmetries of the background FLRW model.
One then finds, e.g., \cite{wass78}:
\beq
\langle B_i(k) B_j^*(k')\rangle = 4\pi^3
\delta^3({k}-{k'}) {P_{ij}(k)} |B(k)|^2\,,
\eeq
where, due to the ${\rm div} {\bf B} = 0$ constraint, $P_{ij}(k)$ must be the 
transverse projection tensor: 
\beq P_{ij}({k}) = \delta_{ij} - \frac{k_i 
k_j}{k^2}\,.  
\eeq 
Assuming the spectrum $B(k)$ is known, then constraints 
at small scales can be used to normalize the spectrum and lead to 
predictions on large scales.

The energy in the magnetic field in a logarithmic k-space 
interval $d\ln k$ is
\beq
\rho_{B}=\frac{d\rho_{B}}{d\ln k} = 
\frac{|B(k)|^2}{8\pi}\,. 
\label{energy}
\eeq 
The evolution of magnetic fields is usually described as 
$B(k) \propto a^{-2}$, which means that $\rho_B$ behaves as isotropic 
radiation.

\section{A simple but effective analytical model}\label{toy}

As we shall see, a most efficient and  elegant amplification mechanism
is to assume a complex scalar field, $\sigma$, charged under $U(1)$, in
addition to the inflaton. We will  assume that its potential,
$V(\sigma\sigma^*)$, is such
that during inflation it is displaced from its global minimum. 
This is relatively easy to arrange and occurs rather naturally
in hybrid models of inflation \cite{LR99} 
\footnote{For example, consider the archetypal 
potential \cite{DSS}
\begin{eqnarray}
V = \alpha^2\phi^2 \sigma^2 +
|\alpha\sigma\sigma^* - \mu^2|^2\,, \nonumber 
\end{eqnarray}
where $\alpha, \mu$ are constants. 
Inflation occurs at $\phi > \phi_{{\rm cr}} \equiv \mu/\alpha$ 
where the minimum of the potential is $\chi = \chi^* = 0$ and hence 
the effective mass of
the photon is zero and the $U(1)$ of electromagnetism is unbroken. For
$A_{\mu} = 0$ and $\phi < \phi_{{\rm cr}}$ the minimum of the potential now
corresponds to the globally SUSY vacuum at $\phi = 0$, $\chi = \chi^* =
\mu/\sqrt{\alpha}$.}.

One way to achieve the desired displacement from
the global minimum is to give $\sigma$ a negative effective mass during
inflation which drives it to  a non-zero vacuum expectation value (vev).
At the end of inflation the effective mass becomes positive and the field
begins coherent oscillations. This is a typical scenario for Affleck-Dine
baryogenesis \cite{Adine} where the coherent oscillations lead to 
the baryogenesis \footnote{A specific model is given by the following 
potential in the supersymmetric standard model (SSM) along a flat direction
\begin{eqnarray}
V = (m_{\sigma}^2 - c H^2)|\sigma|^2 + \left[\ts{\lambda(Am_{3/2} -
aH)\sigma^n\over M^{n-3}} + {\rm h.c.} \right] +
\ts{|\lambda|^2|\sigma|^{2n-6}\over M^{2n-6}}, \nonumber
\label{ssm}
\end{eqnarray}
where $m_{\sigma}$ is of order the weak scale, $m_{3/2}$ is the gravitino
mass and $n$ is proportional to the number of chiral superfields defining
the flat direction. During inflation the $c H^2$ term dominates and drives
$\sigma$ away from the origin. After inflation $\sigma$ oscillates
around the time-dependent minimum of the potential. The terms proportional
to $\lambda$ are soft-supersymmetry-breaking corrections responsible for
violating $B-L$ and giving rise to baryogenesis\cite{DRS}.}.

Giving $\sigma$ a non-zero vev during inflation
spontaneously breaks the $U(1)$ of electromagnetism and causes any
monopole--anti-monopole pairs to be connected by magnetic flux tubes.
These confining flux tubes facilitate the annihilation of monopoles. This,
the Langacker-Pi solution to the monopole problem, is an
independent benefit of breaking conformal invariance in this manner. Such
an additional weapon may be required to deal with monopoles produced by
non-thermal symmetry restoration  in preheating \cite{ntsr} or in models
of inflation which do not solve the monopole problem, such as canonical
$SU(5)$ where the inflaton is a gauge singlet \cite{KDL}.

We will not, however, proceed any further in building a detailed
phenomenology for $\sigma$ but will assume for pedagogical reasons, to
become clear later on, that around the
global minimum the potential is quartic and the field is conformally
coupled to the curvature. The Lagrangian for this scalar QED is:
\beqn
{\cal L}&=&\frac{R}{16\pi G}-D_\mu\sigma(D^\mu \sigma)^*
-\frac{1}{4}F_{\mu\nu}F^{\mu\nu} \nonumber\\
&-&\frac{\lambda_{\sigma}}{4} (\sigma\sigma^*)^2 - 
\frac{1}{12}R |\sigma|^2+{\cal L}_{{\rm inflaton}}\,.
\label{totallag}
\eeqn
The conformal coupling will simplify the evolution equation for $\sigma$
and reduce it to a form independent of $a$. The gauge covariant derivative
$D_{\mu} \equiv \nabla_{\mu} -
i e A_{\mu}$ leads to an effective mass for the photon 
$m_{\gamma}^2 =2e^2|\sigma|^2$ which oscillates in time as $\sigma$ oscillates. This leads
to parametric resonant amplification of $A_{\mu}$ analogous to studies in
Minkowski spacetime.

We work in the so-called unitarity gauge  in which $\sigma = \sigma^*$,
and decompose $\sigma$ into a homogeneous part and a fluctuation:
$\sigma(t, {\bf x})  \to \sigma(t)+\delta\sigma (t,{\bf x})$.
Now let $\sigma(t_i)$ be the initial amplitude of $\sigma$ oscillations.
We assume that the oscillations are independent of the inflaton, $\phi$,
and follow the notation of \cite{BV99} in denoting variables 
rescaled by the scale factor $a$ with a tilde; e.g.,
 $\tilde{\sigma} \equiv a \sigma$.
Then the equation for $\tsig(t)$ is 
\beq
\tsig'' + \lambda_{\sigma} \tsig^3  + 
e^2 a^2 \langle A^2 \rangle \tsig = 0,
\label{sigma}
\eeq
where a prime denotes the derivative with respect to the 
conformal time, $\eta$, and $\langle A^2 \rangle$ is the 
expectation value of  $A_{\mu}A^{\mu}$. 
The electromagnetic field vanishes in 
the background, and hence it is automatically gauge invariant in the 
perturbed spacetime.  Neglecting the last term in Eq.~(\ref{sigma}) for 
the moment and  introducing the dimensionless quantities
$x=\sqrt{\lambda_{\sigma}}\tilde{\sigma}(t_i) \eta$ and 
$f=\tilde{\sigma}(t)/\tilde{\sigma}(t_i)$, we find \beq 
\frac{d^2f}{dx^2}+f^3=0.
\label{f}
\eeq
The solution for this equation can be written as an elliptic cosine,
$f={\rm cn}(x, 1/\sqrt{2})$, 
which yields \cite{GKLS97,dk2} 
\beq \tilde{\sigma} = \tilde{\sigma}(t_i) {\rm cn}\left(x, 
\frac{1}{\sqrt{2}} \right)\,.
\label{cn}
\eeq
The Fourier modes of $\tilde{\sigma}$ fluctuations satisfy
the following equation:
\beq
\frac{d^2}{dx^2} \delta \tilde{\sigma}_k + 
\left[\kappa^2 + 3 {\rm cn}^2 \left(x, \frac{1}{\sqrt{2}} \right)
\right] \delta \tilde{\sigma}_k = 0\,,
\label{delsig}
\eeq
where $\kappa^2=k^2/(\lambda_{\sigma} \tilde{\sigma}^2(t_i))$.

\subsection{Parametric amplification of magnetic fields} 

Variations of the Lagrangian (\ref{totallag}) with respect to $A_{\mu}$
leads to the following equation,
\beq
\nabla_{\mu} F^{\mu\nu}=-j^{\nu}+2e^2A^{\nu}|\sigma|^2,
\label{maxwell2}
\eeq
where the current $j^{\nu}$ is defined by 
$j^{\nu}=ie(\sigma \nabla^{\nu}\sigma^{*}-\sigma^{*} \nabla^{\nu}\sigma)$, 
and vanishes when $\sigma=\sigma^*$.  Adopting the Coulomb or radiation 
gauge conditions, $A_0 = 0, \partial^i A_i = 0$, Fourier modes of $A_i$ 
satisfy \cite{TW,FG} 
\beq 
A_{k}'' + (k^2 + 2e^2a^2\sigma^2)A_{k}=0.
\label{amu}
\eeq
Substituting the solution (\ref{cn}) for  (\ref{amu}), we find
\beq
\frac{d^2}{dx^2}A_{k}+\left[\kappa^2 + \frac{2e^2}{\lambda_{\sigma}} 
{\rm cn}^2 \left(x, \frac{1}{\sqrt{2}}\right) \right] A_{k}=0\,.
\label{amu2}
\eeq
\begin{figure}
\epsfxsize = 3.0in
\epsffile{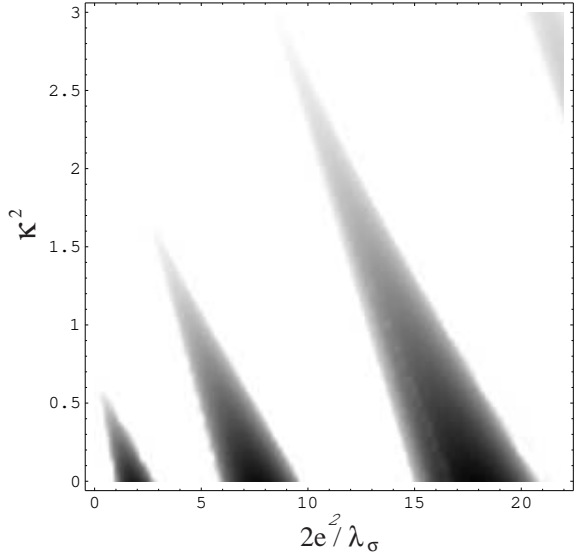}
\caption{Density plot of the Floquet chart for the 
generalized Lam\'e equation (\ref{amu2}) 
for $0 \leq 2e^2/\lambda_{\sigma} \leq 22$ and $\kappa^2\leq 3$.
The shaded regions correspond to parameter ranges where parametric 
amplification of magnetic fields can be expected, $\mu_k > 0$.  
The Floquet index, $\mu_k$, takes larger values in the 
darker shaded regions, and reaches its maxima  for 
$2e^2/\lambda_{\sigma}=2n^2$ at $\kappa^2=0$.
} 
\label{ferminfi}
\end{figure}
\begin{figure}
\epsfxsize = 3.0in
\epsffile{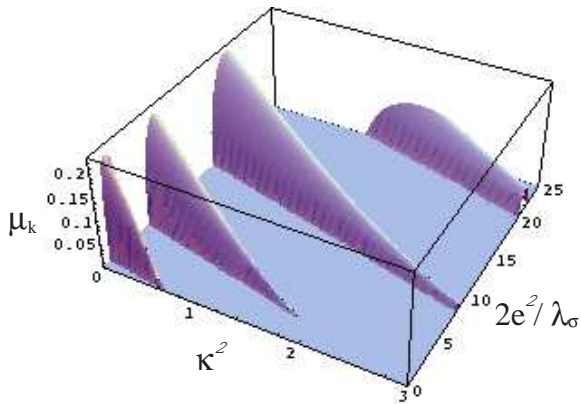}
\caption{Three-dimensional plot of the Floquet chart for the 
generalized Lam\'e equation (\ref{amu2}) 
for $0 \leq 2e^2/\lambda_{\sigma} \leq 25$ and $\kappa^2\leq 3$. 
} 
\label{lame3d}
\end{figure}
The whole system reduces to a problem in Minkowski spacetime 
and hence can be solved exactly using the Floquet theory. 
In fact Eqs.~(\ref{delsig}) and (\ref{amu2}) are the Lam\'e and generalized 
Lam\'e equations respectively.  This elegant exact solution 
is unstable to perturbations 
which introduce a length scale into the problem 
(such as giving $\sigma$ a mass) but the 
existence of the parametric resonance is stable.

The solutions of these equations behave as $\sim e^{\mu_k x}$ where $\mu_k$ is
the Floquet index, which controls the strength of the exponential growth.
As for the solutions of the $\delta\sigma_k$ fluctuation, 
Eq.~(\ref{delsig}), there is only a single resonance band \cite{on}, 
constrained to lie in the narrow, 
sub-Hubble range \cite{GKLS97,dk2,lame}, 
\beq 
\frac32<\kappa^2<\sqrt{3},
\label{deltasig}
\eeq
with a small maximum growth rate, 
$\mu_{\rm max} \approx 0.03598$ at $\kappa^2 \approx 1.615$.
In the absence of the $\sigma$ decay to the magnetic field, 
resonance ends before the energy of the homogeneous $\sigma$ is 
sufficiently transferred to the $\sigma$ fluctuation, in which case 
the final variance is estimated as
 $\langle\delta\sigma^2\rangle \approx 0.05\sigma^2$.

\begin{figure}
\epsfxsize = 3.5in
\epsffile{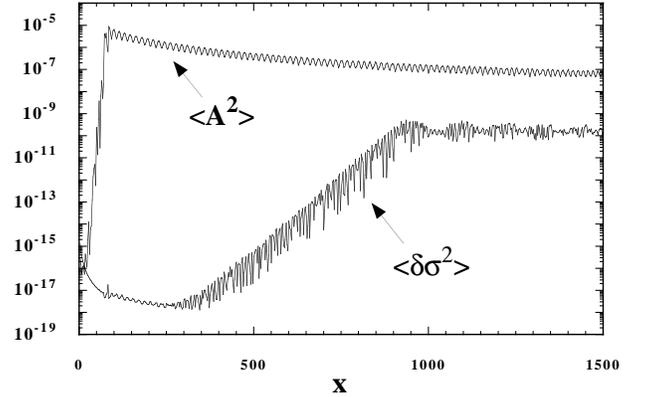}
\caption{The evolution of the variance $\langle A^2 \rangle=
\langle A_{\mu}A^{\mu} \rangle$ and $\langle \delta \sigma^2 \rangle$ 
for $2e^2/\lambda_{\sigma}=5000$ 
in the Hartree approximation at zero conductivity, 
c.f. Fig.~\ref{backnew}.
In this case fluctuations of the magnetic field are dominated
by sub-Hubble modes, and the growth of $\langle A^2 \rangle$
stops by backreaction effects. Note that $\langle \delta \sigma^2 \rangle$ 
is also amplified with smaller growth rate relative to that of 
$\langle A^2 \rangle$.}
\label{backreaction}
\end{figure}

In contrast the magnetic fluctuations can exhibit strong amplifications,
whose strength depends on the ratio, $2e^2/\lambda_{\sigma}$.
According to the analytic investigation in Ref.~\cite{GKLS97}, 
the strongest resonance occurs at $\kappa^2=0$ with 
$\mu_{\rm max}=0.2377$ when the 
parameter $2e^2/\lambda_{\sigma}$ equals 
\beq 
2e^2/\lambda_{\sigma}=2n^2\,,
\label{strongest}
\eeq
where $n$ is an integer. Fluctuations with low momenta
($\kappa \to 0$) are enhanced in the parameter range,
\beq
n(2n-1)<2e^2/\lambda_{\sigma}<n(2n+1)\,,
\label{strong}
\eeq
in which case $\mu_k$ is typically large and strong resonance 
can be expected. 
This is found in Fig.~\ref{ferminfi} where we show a density chart of 
the Floquet index vs $\kappa^2$ and $2e^2/\lambda_{\sigma}$. 

When $2e^{2}/\lambda_\sigma \sim {\cal O}(1)$ the magnetic
field is not suppressed on super-Hubble scales
during inflation \cite{BV99,FB,TBV,ZBS}.  
Since the resonance bands (where 
$\mu_k > 0$) stretch down to include arbitrarily small $k/aH$ in the 
parameter  regions given by (\ref{strong}), 
this allows the resonant production of large-scale, 
coherent, magnetic fields during preheating without violation of causality 
\cite{BKM,BTKM} for the case of $1<2e^2/\lambda_{\sigma}<3$ and 
$6<2e^2/\lambda_{\sigma}<10$.

In Fig.~\ref{backnew} we plot the evolution of $A_k$
for $2e^2/\lambda_{\sigma}=2$ and  a super-Hubble 
mode $\kappa=10^{-25}$. We find that $A_k$ is amplified about  
$10^9$ times by parametric resonance, in which case 
the resultant cosmological magnetic field is large.
However, as we discuss in the next subsection, the growth of conductivity
during preheating and thermalisation counteracts this resonant growth, 
and can overwhelm it completely.

For large $2e^{2}/\lambda_\sigma (\gg 1)$, the inflationary suppression is 
strong \cite{supp}, which makes the large-scale magnetic fields 
negligibly small even if they are amplified by parametric resonance.
This is actually preferable since development of a strong, coherent
magnetic field on cosmological scales would destroy the isotropy of the
background geometry set-up during inflation.
The magnetic spectrum is blue and steep ($\propto k^{3}$) so that 
the variance is dominated by sub-Hubble modes.

Since the magnetic field modes are growing exponentially, 
backreaction effects become important after the fluctuations 
are sufficiently amplified.  
Taking this into account via the one-loop Hartree approximation, 
Eq.~(\ref{sigma}) is modified to
\beq 
\tsig'' + \lambda_{\sigma} (\tsig^2 + 
3\langle \delta\tilde{\sigma}^2 \rangle)\tilde{\sigma}+ 
e^2 a^2 \langle A^2 \rangle \tsig = 0.
\label{sigma2}
\eeq
As long as the ratio $2e^2/\lambda_{\sigma}$ lies in the range of 
Eq.~(\ref{strong}), the growth rate of $A^i_k$ is typically larger than 
that of 
the $\delta\sigma_k$ fluctuation.\footnote{Note that in the limit of 
$2e^2/\lambda_{\sigma} \to \infty$, the maximal $\mu_k$ asymptotically 
approaches the value $0.2377$ for arbitrary $2e^2/\lambda_{\sigma}$ 
\cite{GKLS97}.} This makes to stop the growth of the magnetic field modes 
earlier by backreaction effects when the term $e^2 a^2\langle A^2 
\rangle \tsig$ in Eq.~(\ref{sigma2}) is comparable to the $\lambda_{\sigma} 
\tilde{\sigma}^3$ term, which yields
\beq 
\langle A^2 \rangle \approx \sigma^2/(e^2/\lambda_{\sigma}).
\label{fvariance}
\eeq
This relation indicates that the final variance is suppressed with 
$2e^2/\lambda_{\sigma}$ being increased, which is similar to the standard 
picture of preheating. 
In actual numerical simulations based on the Hartree approximation, 
the final variance typically takes larger values than estimated 
by Eq.~(\ref{fvariance}).  The backreaction effect due to the growth of 
magnetic fluctuations do not completely violate the $\sigma$ oscillations 
\cite{BV99}, which can lead to the amplification of the $\sigma$ 
fluctuations even after magnetic fluctuations are sufficiently amplified.  
This behaviour is found in Fig.~\ref{backreaction} where we plot the 
evolution of fluctuations for the case of $2e^2/\lambda_{\sigma}=5000$.

\subsection{The growth of conductivity, $\sigma_c$} 

The above analysis assumed that the conductivity $\sigma_c$ 
of the universe vanished
during inflation and preheating. This is almost certainly incorrect \cite{GS}
but accurate modelling of the growth of conductivity is difficult for two 
reasons:

(1) such calculations depend sensitively on the underlying 
theory in which the inflaton is embedded, and 

(2) to estimate the growth of conductivity $\sigma_c$ requires 
non-perturbative, non-equilibrium quantum field theory techniques, hence is
extremely difficult. 

(3) Accurate estimates of the final magnetic field requires 
the conductivity in three phases - during inflation, during 
the initial resonance phase and during thermalisation, each of which is 
dominated by different physics. 

The growth of $\sigma_c$ in the QED case has been studied in detail \cite{BDS}.
Since this is not appropriate for energies near the GUT scale we can only 
draw broad lessons: the conductivity grows exponentially
but is also spatially inhomogeneous. 

\begin{figure}
\epsfxsize = 3.2in
\epsffile{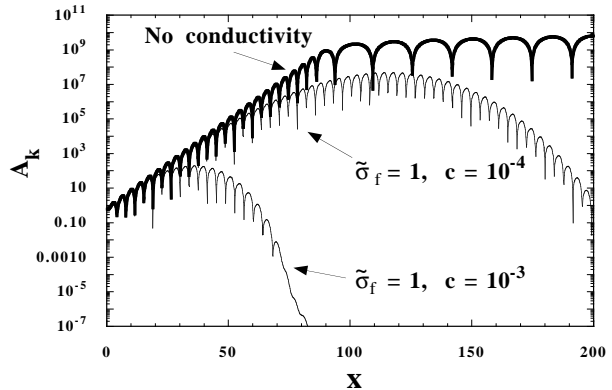}
\caption{The growth of $A_k$ in the case of 
$2e^2/\lambda_{\sigma}=2$ for the mode $\kappa=10^{-25}$.
In the absence of conductivity ($\sigma_c = 0$)
the large-scale magnetic field fluctuations are strongly amplified due 
to the oscillating $\sigma$ field. As the rapidity of the growth of 
the conductivity increases (increasing $c$) the resonant growth of $A_k$
is increasingly stalled. For large conductivity (not shown)  
$\sigma_c \rightarrow \infty$ the electric field vanishes 
and the magnetic field becomes frozen, $B_k \propto
a^{-2}$. }
\label{backnew}
\end{figure}

This is related to the fact that while the plasma is on average charge neutral,
there will be fluctuations in the charge density which act as stochastic 
sources of magnetic fields; see \cite{calzetta}. Given the problems described 
above we take a phenomenological approach to the growth of conductivity. 

Since we are interested in large scales we neglect the spatial variation 
of  $\sigma_c$ and model its growth as:
\begin{equation} 
\sigma_c =  \sigma_f \tanh[c (x-x_0)],
\label{conductivity}
\end{equation} 
where $\sigma_f$ is the final value of conductivity, 
and $c$ controls  the growth rate of $\sigma_c$, $x$ is the dimensionless
conformal time and $x_0$ is the onset of preheating. We therefore assume
that $\sigma_c = 0$ during inflation. As noted in \cite{GS} if 
$\sigma_c > e^2 \sigma^2$ at the onset of preheating then the resonance 
in $A_k$ never begins. 

$\sigma_f$ and $c$ determine the strength of conductivity.
Fig.~\ref{backnew} shows the evolution of a cosmological $A_k$ mode 
for three pairs $(\sigma_f, c)$. The value $\tilde{\sigma}_f \equiv
\sigma_f/(\sqrt{\lambda_{\sigma}}\sigma(t_i))=1$ is used, where
$\sigma(t_i)$ is the value of the $\sigma$ field at the beginning 
of its oscillations.

The four-potential in the finite-$\sigma_c$ case obeys the equation:
\begin{eqnarray} 
A_k''  + (k^2+2e^2a^2\sigma^2) A_k =-\sigma_c a A_k',
\label{acond}
\end{eqnarray} 
which shows how the conductivity acts to damp the resonance when we neglect
the spatial dependence of $\sigma_c$. Fig.~\ref{backnew} shows how the
the preheating resonance competes with the damping due to conductivity. 
If the growth of $\sigma_c$ is too rapid (large $c$) the resonant growth 
of $A_k$ is stalled. However, for relatively slow growth of $\sigma_c$ 
(roughly more than 10 $\sigma$ oscillations) $A_k$ can grow almost to its 
$\sigma_c = 0$ maximum. 

Once backreaction causes the resonance to end there is nothing to 
compensate the damping effects of the finite conductivity and $A_k$ begins 
to decay exponentially.  In the limit $\sigma_c \rightarrow 
\infty$ the solution of Eq.~(\ref{acond}) is obviously 
$A_k = {\rm constant}$, which corresponds to 
$B_k \propto a^{-2}$, the ratio of  magnetic field  energy density to 
incoherent radiation energy density is fixed. 

To estimate $r$ at the start of the radiation dominated phase is therefore 
a subtle issue because one must know, not only the growth of conductivity
during the initial preheating phase, but also how the conductivity 
grows during thermalisation and how $\sigma$ decays by Born process to 
complete reheating.  If the conductivity is high during preheating, the 
magnetic fields will exhibit exponential suppression during which 
$\sigma_c$ increases from zero to the final value, $\sigma_f$, which means 
that the gains of preheating will be washed out and lost.

When the conductivity term is much smaller than the $2e^2a^2\sigma^2 A_k$ 
term in Eq.~(\ref{acond}) during preheating, the evolution of magnetic fields
is the same as the case of the non-conductivity in preheating phase.
However when the rhs of Eq.~(\ref{acond}) becomes of order the
$2e^2a^2\sigma^2 A_k$ term after preheating, the magnetic field begins to be 
exponentially suppressed.  

Although the $A_k$ freeze when the lhs of 
Eq.~(\ref{acond}) becomes negligible relative to the conductivity term,
the gains obtained in preheating are not generally preserved due to 
the rapid decay of magnetic fields before the freeze of $A_k$. 
However, the Born decay of $\sigma$ before thermalisation can alter
the strength of the $2e^2a^2\sigma^2 A_k$ term, which may alter the above 
estimates.  In addition to this, we need to know the evolution of 
conductivity during thermalization for a complete study, although it is 
difficult and few studies of thermalization after preheating exist;
see e.g. \cite{son}.

\begin{figure}
\epsfxsize = 3.2in
\epsffile{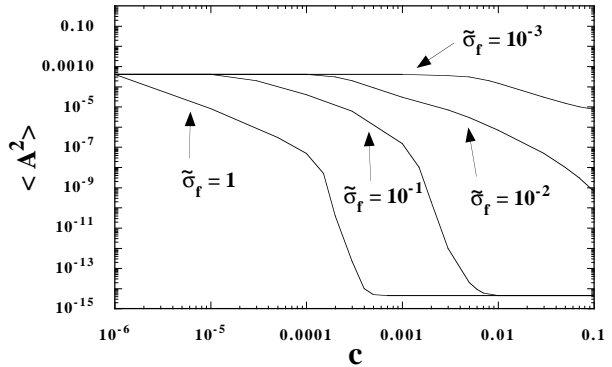}
\caption{The maximum variance $\langle A^2 \rangle$ as a function of 
the conductivity parameters $c$ and $\tilde{\sigma}_f$ appearing in Eq. 
(\ref{conductivity}). As $c$ and $\tilde{\sigma}_f$ increase the period 
and degree to which preheating can amplify  $\langle A^2 \rangle$ decreases
rapidly.}
\label{conductfig}
\end{figure}

In conclusion, in the absence of conductivity,
either the magnetic field is resonantly amplified on super-Hubble scales, 
or it has a $k^3$ spectrum and is too small on cosmological scales.

When conductivity is included one introduces several model-dependent 
parameters into the problem which  impact on the viability of 
preheating as a significant source of magnetic fields. 
The effect of resonance is typically washed out by the growth of 
conductivity, while the final size of magnetic 
fields depends on details of the 
Born decay process and the evolution of conductivity after preheating.
To answer which case applies is model-dependent and beyond the scope of the 
current paper.

\section{Geometric magnetisation}

Turner and Widrow \cite{TW} found that the most efficient way to produce 
magnetic fields is to break $U(1)$ gauge invariance as well as conformal 
invariance, via the Lagrangian: 
\begin{eqnarray} 
{\cal L} &=& \frac{R}{16\pi G} - \frac{1}{4} F_{\mu\nu}F^{\mu\nu}
-\frac{\beta}{2} R A_{\mu} A^{\mu} \nonumber\\
 &-& \frac{\gamma}{2} R_{\mu \nu} A^{\mu} A^{\nu} 
 + {\cal L}_{{\rm inflaton}},
 \label{laggeometric}
\end{eqnarray} 
where $\beta, \gamma$ are real constants.  These terms were 
found to give rise, for apparently reasonable values of the constants, 
during inflation, to fields corresponding to the required 
value $r \sim 10^{-8}$ today \cite{TW}.
Variation of the action (\ref{laggeometric}) with respect to
$A^{\mu}$ and $F^{\mu \nu}$ yield the following equations of motion:
\begin{eqnarray} 
\nabla^{\mu}F_{\mu \nu}-\beta RA_{\nu}-\gamma R^{\mu}_{\nu}
A_{\mu}=0,
\label{max1}
\end{eqnarray} 
\begin{eqnarray} 
\partial_{\mu}F_{\nu \lambda}+\partial_{\lambda}F_{\mu \nu}
+\partial_{\nu}F_{\lambda \mu}=0.
\label{max2}
\end{eqnarray} 
Writing these equations in terms of magnetic and electric fields 
and eliminating the electric field, we obtain \cite{TW}
\begin{eqnarray} 
(a^2 {\bf B})''-\nabla^2 (a^2{\bf B})+\theta(\eta) a^2{\bf B}=0,
\label{geobasic1}
\end{eqnarray} 
with 
\begin{eqnarray} 
\theta(\eta)=6\beta \frac{a''}{a}+\gamma \left\{\frac{a''}{a}+ 
\left(\frac{a'}{a}\right)^2 \right\}.
\label{theta}
\end{eqnarray} 
Expanding the magnetic field in Fourier components as
$a^2{\bf B}=\int e^{-i{\bf k}\cdot{\bf x}}\cb~d^3{\bf k}$, each 
mode satisfies the 
following equation,
\begin{eqnarray} 
\cb''+\left[k^2+\theta(\eta)\right]\cb=0.
\label{Feq}
\end{eqnarray} 
Hereafter we set $\gamma=0$ and leave $\beta$ 
as a free parameter, in which case $\theta$ reduces to
$\theta=6\beta a''/a=6\beta a^2 R$.
When the system is dominated by the inflaton field, $\phi$,
the scalar curvature is:
\begin{eqnarray} 
R=\frac{8\pi}{m_{\rm pl}^2}
 \left[4V(\phi)- \frac{\phi'^2}{a^2} \right],
\label{curvature}
\end{eqnarray} 
where $V(\phi)$ is the inflaton potential.
During inflation $R$ slowly decreases.  
When $\beta$ is negative, the magnetic field 
fluctuations exhibit super-adiabatic amplification due to the so-called
negative coupling instability, as studied in the non-minimally coupled 
case in Refs.~\cite{SH,TY}. This enhancement is most relevant during 
inflation. 

\subsection{Magnetic amplification during inflation}

Let us study the evolution of magnetic fluctuations during inflation.
When $\gamma=0$, the solutions for Eq.~(\ref{Feq}) are 
expressed as combinations of the Hankel
functions $H_{\nu}^{(J)}$ ($J=1, 2$) \cite{SH}:
\begin{eqnarray}
\cb=c_1 \sqrt{\eta} H_{\nu}^{(2)}
(k \eta) +c_2 \sqrt{\eta} H_{\nu}^{(1)}(k \eta),
\label{analyticFk}
\end{eqnarray}
where $c_1$ and $c_2$ are constants, and the order $\nu$ of
the Hankel functions is given by 
\begin{eqnarray}
\nu^2=\frac14 -12\beta.
\label{nu}
\end{eqnarray}
The choice of $c_1=\sqrt{\pi}/2$ and $c_2=0$ corresponds to
the Bunch-Davies vacuum.
In the long wavelength limit, $k\eta \to 0$, 
$H_{\nu}^{(2,1)}(k\eta)$ approaches the values
\begin{eqnarray}
H_{\nu}^{(2,1)}(k\eta) \to \pm \frac{i}{\pi} \Gamma(\nu)
\left(\frac{k\eta}{2}\right)^{-\nu}, 
\label{longlimit}
\end{eqnarray}
where $\Gamma(\nu)$ is the Gamma function.
In inflation, conformal time can approximately be
written as $\eta \approx -1/(aH)$, and  
long wave $\cb$ modes exhibit exponential growth, 
\begin{eqnarray}
\cb \propto a^{\nu- 1/2}=a^{\frac12(\sqrt{1-48\beta}-1)},
\label{Fkgrowth}
\end{eqnarray}
for negative values of $\beta$.
In this case the energy density in the $k$-th mode of the magnetic 
field evolves as
\begin{eqnarray}
\rho_B \propto |\cb|^2/a^4=a^{\sqrt{1-48\beta}-5},
\label{rhob}
\end{eqnarray}
which means that the ratio, $r=\rho_B/\rho_{\gamma}$, {\em increases} during 
inflation when $\beta<0$.
This makes it possible to reach the value $r \sim 10^{-8}$
required to explain the existence of current galactic magnetic
fields \cite{TW}.

Large negative values of $\beta$ lead to extremely strong amplification 
of magnetic fields.  When $\beta=-1/6$, 
$\cb \propto a$ and $\rho_B \propto a^{-2}$, which 
corresponds to the minimally coupled scalar field case.
Compared with the standard adiabatic result, $\rho_B 
\propto a^{-4}$ with $\beta=0$, the energy density decreases
more slowly due to superadiabatic amplification.

For $\beta ~<~1$, super-Hubble magnetic fluctuations 
exhibit enormous amplification during inflation, i,e., 
$r=\rho_B/\rho_{\gamma} \propto a^c$ with $c \ge 6$, 
which conflicts with  observations unless their initial values at the start
of inflation were extraordinarily small. 

When $\beta$ is positive, magnetic fields are exponentially 
suppressed during inflation.
For $0<\beta<1/48$, which corresponds to $0<\nu<1/2$,
$\cb$ and $\rho_B$ evolve as
\begin{eqnarray}
\cb \propto a^{-\frac12(\sqrt{1-48\beta}-1)},~~~~
\rho_B \propto a^{-\sqrt{1-48\beta}-3}.
\label{positive1}
\end{eqnarray}
When $\beta>1/48$ (i.e., complex $\nu$), we find 
\begin{eqnarray}
\cb \propto a^{-1/2},~~~~
\rho_B \propto a^{-5},
\label{positive2}
\end{eqnarray}
in which case the evolution of magnetic fields is independent of 
the strength of $\beta$.

\begin{figure}
\epsfxsize = 3.5in
\epsffile{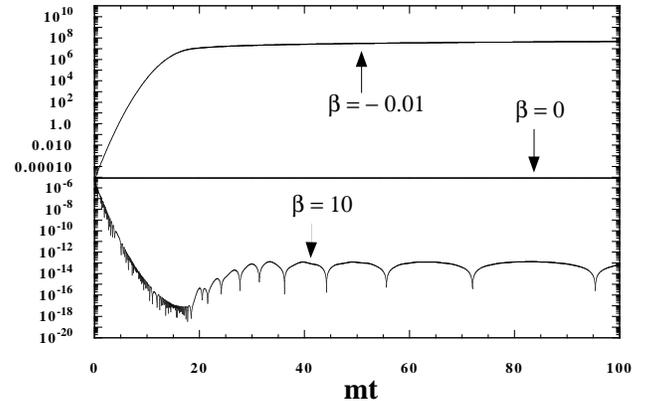}
\caption{The evolution of a super-Hubble $\cb$ mode during inflation 
and preheating for $\beta=0, -0.01, 10$ in the massive chaotic inflationary 
model.  We choose the initial value of inflaton as $\phi=3m_{\rm pl}$,
which corresponds to 60 e-foldings before the end of inflation.
The inflationary period continues until $mt \sim 20$, after which 
the system enters the reheating stage.  While $\cb$ is exponentially 
suppressed during inflation for positive $\beta$, negative $\beta$
leads to superadiabatic amplification.} 
\label{geometric}
\end{figure}

We plot in Fig.~\ref{geometric} the evolution of a super-Hubble $\cb$ mode
for $\beta=0, -0.01, 10$ for the inflaton potential,
\begin{eqnarray}
V(\phi)=\frac12 m^2\phi^2.
\label{geopotential}
\end{eqnarray}
When $\beta=0$, $\cb$ 
is constant (i.e., $B_k \propto a^{-2}$) from Eq.~(\ref{positive1}),  
as is confirmed in Fig.~\ref{geometric}.  For negative $\beta$, $\cb$ exhibits 
an exponential increase as estimated by Eq.~(\ref{Fkgrowth}).  An important 
point to note is that the rapid growth of magnetic fields may 
affect the evolution of background quantities, an effect we do not include.  

In Ref.~\cite{TY}, it was found 
that exponential growth of scalar field fluctuations makes the inflationary 
period terminate earlier, in the context of a non-minimally coupled scalar 
field.  Exponential growth of large scale magnetic fields will also 
stimulate the enhancement of super-Hubble metric perturbations, which may 
lead to deviations from the scale-invariant Harrison-Zel'dovich spectra.  
This is expected to be strong for $\beta~\lsim~-1$ in the analogy of the 
non-minimally coupled scalar field case \cite{TY}.  Complete analysis 
including backreaction and metric perturbations is now in progress.

\subsection{The preheating phase}

The inflationary period corresponds to $0<mt~\lsim~20$ in 
Fig.~\ref{geometric}, after which time the system enters the reheating stage.  
During reheating, 
the scale factor evolves as $a \propto t^{2/3} \propto \eta^2$ in the 
massive inflaton potential (\ref{geopotential}).  From 
Eqs.~(\ref{analyticFk}) and 
(\ref{longlimit}), we have that $\cb \propto 
a^{\frac14(1+\sqrt{1-48\beta})}$ and $\rho_B \propto 
a^{\frac12(\sqrt{1-48\beta}-7)}$ for negative $\beta$ in the long-wave 
limit $k\eta \to 0$.  Similarly $\cb \propto a^{1/4}$ and 
$\rho_B \propto a^{-7/2}$ when $\beta>1/48$.  However, 
this corresponds to an estimate of the frequency $\theta 
\propto \eta^{-2} \propto t^{-2/3}$, which only provides information about 
the average amplitude of the scalar curvature.

In actual fact the scalar curvature oscillates due to the oscillating inflaton 
condensate, which can lead to efficient enhancement of field fluctuations 
\cite{BL98,TMT,TB00}.  We find in Fig.~\ref{geometric} that $\cb$ begins to 
grow for $mt~\gsim~20$ in the case of $\beta=10$ in spite of the 
inflationary suppression.  This is the geometric preheating stage where 
$\cb$ grows quasi-exponentially, in which case the above naive estimate 
neglecting the oscillations of the scalar curvature can not be applied.  In 
the non-minimally coupled multi-field case, the growth of scalar field 
fluctuations during preheating is only relevant for $|\xi|~\gsim~1$ 
\cite{BL98,TMT}.

\begin{figure}
\epsfxsize = 1in
\caption{{\bf See associated JPEG image}. 
The time-averaged Floquet chart for the geometric magnetisation 
mechanism during preheating. Since the problem 
is {\em not} scalar-factor independent, exact 
Floquet theory cannot be used - the expansion causes the weakening 
of the resonance bands but also removes the stability bands so that all 
modes are amplified for sufficiently large $\beta$. It is clear that as $\beta$
is increased, the super-Hubble modes $\kappa \ll 1$ are the ones that grow 
first. The negative coupling case $\beta < 0$, which is much stronger, is not
shown.}
\label{geomfloquet}
\end{figure}

Let us analytically study the evolution of magnetic field fluctuations 
during preheating.  Making use of the time averaged relation, $\langle\frac12 
\dot{\phi}^2\rangle_T=\langle V(\phi) \rangle_T$, with the potential 
(\ref{geopotential}), the evolution of the inflaton condensate 
is described by 
\begin{eqnarray}
\phi=\frac{m_{\rm pl}}{\sqrt{3\pi}mt} \sin mt, 
\label{oscillation}
\end{eqnarray}
where we choose the time when the oscillation starts 
as $mt_0=1/4$ \cite{KLS}.  
Then the scalar curvature (\ref{curvature}) is approximately 
\begin{eqnarray}
R \approx \frac{4}{3t^2}(1-3\cos2mt).
\label{curvature2}
\end{eqnarray}
Although $R$ oscillates, its amplitude
decreases as $t^{-2}$ due to the cosmic expansion
which means that parametric resonance soon becomes ineffective
if $|\beta|$ is small.
Substituting Eq.~(\ref{curvature2}) into Eq.~(\ref{Feq}) and introducing 
a new scalar field $\bar{\cal B}_k=a^{1/2}\cb$, $\bar{\cal B}_k$ satisfies the 
well known Mathieu equation\footnote{We neglect the 
$(\dot{a}/{2a})^2-\ddot{a}/{2a}$ term which appears in the 
parenthesis of Eq.~(\ref{mathieu}), which can be justified for 
$|\beta|~\gsim~1$.}, 
\begin{eqnarray}
\frac{d^2 \bar{\cal B}_k}{d z^2} + \left(A_k -2q \cos 2z \right)
\bar{\cal B}_k=0,
\label{mathieu}
\end{eqnarray}
where for $\beta>0$,
\begin{eqnarray}
A_k= \frac23 q + \frac{k^2}{m^2a^2},~~~~
q=\frac{3\beta}{\pi^2\bar{t}^2},
 \label{Aq1}
\end{eqnarray}
and for $\beta<0$,
\begin{eqnarray}
A_k= -\frac23 q + \frac{k^2}{m^2a^2},~~~~
q=\frac{3|\beta|}{\pi^2\bar{t}^2}.
 \label{Aq2}
\end{eqnarray}
Here $z=mt$, and $\bar{t}=mt/(2\pi)$ naively corresponds to the number of 
oscillations executed by the inflaton at time $t$.

In the context of standard preheating with the effective potential, 
$V(\phi,\chi)=\frac12 m^2\phi^2+\frac12g^2\phi^2\chi^2$,
the relation of $A_k$ and $q$ for the $\chi$ field is written as 
$A_k=2q+k^2/(m^2a^2)$ \cite{KLS}.
In this case $\chi$ particle production is inefficient 
unless the initial $q$ is much larger than unity.
In contrast, the resonance band is broader  in the present model\cite{TMT}.

Therefore parametric resonance takes place for smaller initial values of $q$.
In spite of this, since the resonance band is narrow for $q~\lsim~1$,
we typically require the coupling, $|\beta|~\gsim~1$, for  relevant
growth of $\cb$\footnote{When $|\beta|=1$, the initial value of $q$ is 
estimated to be $q_i \approx 4.9$ with $\bar{t}_i=1/4$.} (see 
Fig.~\ref{geomfloquet} where we show Floquet indices for positive $\beta$).  
When $|\beta|~\gsim~1$, fluctuations grow as $\bar{\cal B}_k \sim e^{\mu_k 
mt}$, whose growth rate gets gradually larger with $|\beta|$.  For $|\beta| 
\gg 1$, however, the final variance of magnetic fields will be suppressed 
as studied in e.g.,  Ref.~\cite{TMT}.

When $\beta$ is positive, long-wave $\cb$ modes 
are exponentially suppressed during inflation. Hence it is rather 
difficult to produce sufficient 
large scale magnetic fields even when $\beta \gg 1$.  On 
sub-Hubble scales, however, the inflationary suppression of $\cb$ is weak 
relative to super-Hubble modes and magnetic fields are excited during 
preheating.  Hence the final magnetic variance 
$\langle A^2 \rangle$, is dominated by sub-Hubble modes.
For negative values of $\beta$ with $\beta~\lsim~-1$, the growth of $\cb$ 
can be strong but is typically dominated by the growth during inflation.  
When $-1~\lsim~\beta<0$, the production of magnetic fields is weak 
during preheating, while they are amplified in the preceding inflationary 
phase as found in Fig.~\ref{geometric}.

\subsection{Effects of the growth of conductivity}

Now let us consider the ratio $r=\rho_B/\rho_\gamma$ on some comoving 
scale in the presence of the preheating phase.
As the reheating process proceeds, the effect of the conducting plasma
is expected to be important \cite{TW}. 
This effect appears as a friction-like term in the equation of $\cb$ 
\begin{eqnarray}
\cb''+\left[k^2+\theta(\eta)\right]\cb=-\sigma_c a \cb'
\label{conduct}
\end{eqnarray}
where $\sigma_c$ is the conductivity of the plasma.
If the conductivity is very high, we find $\cb \sim {\rm constant}$,
which implies that the energy density of magnetic fields decreases
as $\rho_B \sim a^{-4}$.

We assume that the effect of the conductivity begins to dominate
at some temperature, $T_c~(\lsim~T_{r})$, where $T_r$ is 
the reheating temperature.  At the first Hubble-crossing 
during inflation, the ratio of $\rho_B$ to the total energy density, 
$\rho_T$, is approximately 
estimated as $\rho_B/\rho_T \approx (M/m_{\rm pl})^4$, 
where $M^4$ is the energy scale of inflation.  For negative $\beta$, one 
obtains the following ratio on the comoving length scale $\lambda$ 
neglecting the parametric amplification of magnetic fields during 
preheating \cite{TW}: 
\begin{eqnarray}
r &\approx& 10^{26(\alpha-5)} \left( \frac{M}{m_{\rm pl}} 
\right)^{4\alpha/3} \left(\frac{T_r}{m_{\rm pl}} \right)^{(1+3\alpha)/3} 
\nonumber \\
&\times& \left(\frac{T_c}{m_{\rm pl}}
\right)^{-4(1+\alpha)/3} \left(\frac{\lambda}{{\rm Mpc}}\right)
^{\alpha-5},
\label{ratioest}
\end{eqnarray}
where $\alpha \equiv \sqrt{1-48\beta}$.
For example, when $\beta=-1/2$, $r |_{\rm 1Mpc}$ reaches $\sim 10^{-8}$ for 
$M=10^{17}$ GeV, $T_r=10^{17}$ GeV, and $T_c=10^{17}$ GeV, in which case 
seed magnetic fields can be produced without the need for the 
galactic dynamo mechanism.  
When $-1~\lsim~\beta<0$, since parametric excitation of magnetic 
fields is irrelevant, the estimation of $r$ in Eq.~(\ref{ratioest}) is hardly 
modified due to the existence of the preheating phase.  

In contrast, for 
$\beta~\lsim~-1$, it is expected that preheating will lead to the increase 
of $r$.  In this case, however, $r |_{\rm 1Mpc}$ is typically much 
greater than unity 
even in the absence of preheating because $10^{26(\alpha-5)} \gg 1$ in 
Eq.~(\ref{ratioest})\footnote{For example, for $\beta=-1$, $M=10^{16}$ 
GeV, $T_r=10^{15}$ GeV, and $T_c=10^{15}$ GeV, we find 
$r |_{\rm 1Mpc} \sim 10^{37}$, 
which is clearly excessive.}.  Although $r$ is further increased 
corresponding to the amplification of $\cb$ during preheating, this case 
will be ruled out by observations.

Let us consider the case $\beta~\gsim~1$
where the excitation of magnetic fields by resonance is expected.
In this case, an analytic estimate of $r$ neglecting the contribution during 
preheating is 
\begin{eqnarray}
r &\approx& 10^{-130} \left(\frac{M}{m_{\rm pl}}\right)
\left(\frac{T_r}{m_{\rm pl}} \right)^{1/3} \nonumber \\
&\times&
 \left(\frac{T_c}{m_{\rm pl}} \right)^{-4/3} \left(\frac{\lambda}{{\rm 
Mpc}}\right)^{-5}.
\label{ratioest2}
\end{eqnarray}
Due to the strong inflationary suppression, $r$ is restricted to be very small.
For example, for  $M=10^{16}$ GeV, $T_r=10^{15}$ GeV, 
and $T_c=10^{15}$ GeV, $r |_{1{\rm Mpc}}=10^{-129}$.  During preheating, 
the $\cb$ fluctuation exhibits exponential increase, which makes $r$ larger 
than estimated in Eq.~(\ref{ratioest2}).  For example, when $\beta=10$, $\cb$ 
is amplified about $10^5$ times (see Fig.~\ref{geometric}), and the ratio 
increases to $r |_{1{\rm Mpc}} \sim 10^{-120}$.  However, the amplification 
during preheating in the positive $\beta$ case is typically insufficient to 
explain the large-scale seed magnetic fields even for $\beta \gg 1$.

We conclude that with regard to the geometric magnetisation mechanism, 
the ratio $r=\rho_B/\rho_T$ is mainly determined by the 
inflationary phase, despite the fact that  magnetic 
fields can be amplified during preheating. While we have studied this in
the massive inflaton model, we expect similar results in other
inflationary models. For example, in the quartic inflaton potential the 
frequency $\theta$ depends explicitly on the scale factor and we cannot 
reduce the problem to one in Minkowski space, as we did in section III.

\section{Magnetic field amplification due to large metric perturbations}

Since the FLRW metric is conformally flat, i.e., the Weyl tensor vanishes, 
magnetic fields are not produced due to the cosmic expansion.  
During preheating however, scalar metric perturbations can grow 
exponentially on both super-Hubble and sub-Hubble scales \cite{BV99,BTKM}.

This growth of metric perturbations means that the spacetime may 
no-longer be well-described by a conformally flat background metric. If the 
metric perturbations remain small, this breaking of conformal invariance 
is small (as measured by the
curvature invariant 
$C_{\alpha\beta\mu\nu} C^{\alpha\beta\mu\nu}$) and the production of 
photons is very suppressed.  Once the metric perturbations at a certain 
scale become large, however, this is no-longer true and the production of 
magnetic fields can be expected. This was discussed by Calzetta and Kandus
\cite{CK} in the context of structure formation and suggested 
in the context of preheating in \cite{BGMK}. Here we follow the recent 
analysis of Maroto\cite{maroto}.

The  line element for a flat FLRW model with scalar metric perturbations
in the conformal Newtonian or longitudinal 
gauge is \cite{MFB,earlympre}
 \beq ds^2=a^2(\eta)[-(1+2\Phi)d\eta^2 
+(1-2\Phi)\delta_{ij}dx^idx^j].
\label{metric}
\eeq
We consider the following two-field model in the presence of magnetic 
fields:
\begin{eqnarray} 
{\cal L} = \frac{R}{16\pi G} &-& \frac{1}{4} F_{\mu\nu}F^{\mu\nu}
-\frac12 (\nabla \phi)^2-\frac14\lambda\phi^4 \nonumber \\
&-& \frac12 (\nabla \chi)^2 -\frac12g^2\phi^2\chi^2,
\label{lagmetric}
\end{eqnarray} 
where $\chi$ is a scalar field coupled to inflaton, $\phi$.
Then the magnetic field satisfies the Maxwell equation, 
$\nabla_{\mu}F^{\mu\nu}=0$, i.e.,
\begin{eqnarray} 
\frac{\partial}{\partial x^{\mu}}
\left[ \sqrt{-g} g^{\mu \alpha}g^{\nu \beta}
(\partial_{\alpha}A_{\beta}-\partial_{\beta}A_{\alpha})
\right]=0.
\label{maxwell}
\end{eqnarray} 
Using the relations $\sqrt{-g}=a^4(1-2\Phi), 
g^{00}=-a^{-2}(1-2\Phi), g^{ii}=a^{-2}(1+2\Phi)$ in the 
perturbed metric (\ref{metric}), Eq.~(\ref{maxwell}) yields
for $\nu=i$:
\begin{eqnarray} 
& &\frac{\partial}{\partial \eta}\left[(1-2\Phi)
(\partial_i A_0-\partial_0A_i)\right] \nonumber \\ 
&+& 
\frac{\partial}{\partial x^j}\left[(1+2\Phi) 
(\partial_j A_i-\partial_i A_j)\right]=0.
\label{nui}
\end{eqnarray} 
Adopting the Coulomb gauge condition: $A_0=0, \partial^i A_i=0$,
one finds that
\begin{eqnarray} 
A_i''-\nabla^2 A_i &=& 2\Phi' A_i'+4\Phi \nabla^2 A_i \nonumber \\
&+& 2\frac{\partial \Phi}{\partial x^j}
\left(\frac{\partial A_i}{\partial x^j}- 
\frac{\partial A_j}{\partial x^i} \right).
\label{perturbedmax}
\end{eqnarray} 
The effect of metric perturbations appears at second order 
in the rhs of Eq.~(\ref{perturbedmax}). In Fourier space this leads 
to convolutions of the form $\int d^3k' \Phi'_{{\bf k'}} 
A'_{{\bf k}-{\bf k'}}$, which lead to mode-mode coupling.

However, if we assume that $\Phi$ is only dependent on 
time on scales larger than some 
cosmological scale $\lambda_c=2\pi/k_c$ \cite{maroto}, each Fourier 
component of $A_i$ satisfies the simple equation:
\begin{eqnarray} 
A_k''+k^2 A_k=2\Phi'A_k'-4k^2\Phi A_k.
\label{simplify}
\end{eqnarray} 
where the coupling between the metric potential on smaller scales 
($k>k_c$) and the magnetic fields are ignored. Note that one cannot simply
assume $\Phi = \Phi(t)$ on all scales since then the Weyl tensor vanishes 
identically and no photons are produced. 

Treating the full problem is complicated due to the fact that the last 
term in the rhs of Eq.~(\ref{perturbedmax}) does not vanish, hence the various
components of $A_j$ are coupled.  While the 
precise analysis including these fully nonlinear effects is very 
complicated, we can still estimate the amplitude of magnetic fields 
produced during preheating by using Eq.~(\ref{simplify}).

Introducing a new field, $\tilde{A}_{k}=(1-\Phi)A_{k}$, to eliminate
the $A_{k}'$ term in Eq.~(\ref{simplify}), one finds
\begin{eqnarray} 
\tilde{A}_k''+k^2 \tilde{A}_k=\Phi'' \tilde{A}_k,
\label{simplify2}
\end{eqnarray} 
where we have  neglected the last term in Eq.~(\ref{simplify}) which is not 
important on large scales.
Before the start of preheating, the $\Phi''\tilde{A}_{k}$ term 
is negligible
and  $\tilde{A}_{k}$ is described by the following 
positive-frequency solution:
\begin{eqnarray} 
\tilde{A}_{k}^{(i)} \approx \frac{1}{\sqrt{2k}}e^{-ik\eta}.
\label{solutionini}
\end{eqnarray} 
One finds the solution for Eq.~(\ref{simplify2}) in integral form 
\cite{ZS}:
\begin{eqnarray} 
\tilde{A}_k(\eta)=\tilde{A}_k^{(i)}+ \frac{1}{k} \int_{\eta_i}^{\eta} 
\Phi'' \tilde{A}_k (\eta') \sin k(\eta-\eta') d\eta'.
\label{solutioneta}
\end{eqnarray} 
The energy density in the magnetic field can be expressed as
\begin{eqnarray} 
\rho_B=(k/a)^4 |\beta_k|^2,
\label{energyden}
\end{eqnarray} 
where the Bogolyubov coefficients, $\beta_k$, are approximately \cite{maroto}
\begin{eqnarray} 
\beta_k=-i \int_{\eta_i}^{\eta}
\tilde{A}_k^{(i)}\Phi'' \tilde{A}_k(\eta) d\eta.
\label{beta}
\end{eqnarray} 
Substituting Eq.~(\ref{solutioneta}) for Eq.~(\ref{beta}) with 
Eq.~(\ref{solutionini}) and assuming that $\Phi'$ vanishes 
before and after preheating [i.e., $\Phi'(t_i)=0$ and $\Phi'(t_f)=0$
where the subscript $i$ and $f$ denote the values at the beginning and end
of preheating, respectively], one easily finds that the next order term in 
Eq.~(\ref{solutioneta}) gives an important contribution to $\beta_k$, 
yielding \cite{maroto}
\begin{eqnarray} 
\beta_k \approx -\frac{i}{2k}
\int_{\eta_i}^{\eta_f} (\Phi')^2 d\eta,
\label{beta2}
\end{eqnarray} 
where we considered the super-Hubble modes: $k\eta \ll 1$.
Combining Eqs.~(\ref{energy}), (\ref{energyden}), and (\ref{beta2}),
we obtain the amplitude of magnetic fields as
\begin{eqnarray} 
|B_k| \approx \frac{k}{a^2} \int_{\eta_i}^{\eta_f}
(\Phi')^2 d\eta.
\label{magestimation}
\end{eqnarray}  
In order to analyze the evolution of the magnetic fields, it is convenient to 
rewrite Eq.~(\ref{magestimation}) using the dimensionless conformal time, 
$x =\sqrt{\lambda}a_i\phi_i \eta$, as
\begin{eqnarray} 
|B_k|=\left(\frac{a_i}{a}\right) \left(\frac{k}{a}\right)
\sqrt{\lambda}\phi_i \int_{x_i}^{x_f}
\left(\frac{d\Phi}{dx}\right)^2dx.
\label{magestimation1}
\end{eqnarray} 
At the decoupling epoch where the coherence scale corresponds
to $(k/a)_{\rm dec} \sim 10^{-33}$ GeV, the amplitude of 
magnetic fields can be estimated by
\begin{eqnarray} 
|B_k^{\rm dec}|/ 1 {\rm G} \approx 10^{-2} \frac{a_i}{a_{\rm dec}} 
\int_{x_i}^{x_f}
\left(\frac{d\Phi}{dx}\right)^2dx,
\label{magestimation2}
\end{eqnarray} 
where we used the value, $\sqrt{\lambda}\phi_i \sim 10^{12}$ GeV.
The ratio $a_i/a_{\rm dec}$ depends on the reheating temperature, $T_R$.
If the energy of inflaton at the end of inflation were instantaneously  
transferred to radiation, the reheating temperature would be 
$T_R \sim 10^{15}$ GeV, which yields $a_i/a_{\rm dec}\sim T_{\rm 
dec}/T_R~\sim~10^{-25}$.  Note that the ratio $a_i/a_{\rm dec}$ becomes 
larger for lower reheating temperature.

Primordial seed magnetic fields for the galactic dynamo mechanism are in 
the regions of $|B_k^{\rm dec}|=10^{-25}$ G $\sim 10^{-15}$ G.
In the single field 
case in which large-scale metric perturbations are hardly amplified 
during preheating, 
it was found that magnetic fields estimated by Eq.~(\ref{magestimation2}) 
are below the values required for the galactic dynamo in the realistic
values of $a_i/a_{\rm dec}$ \cite{maroto}.

In the two-field case with self-coupling inflaton, we can expect the growth 
of metric perturbations due to the enhancement of field perturbations, 
which stimulates the growth of magnetic fluctuations through gravitational 
scattering.
Decomposing the scalar fields as $\varphi_J(t, {\bf x}) \to \varphi_J(t)+
\delta\varphi_J(t, {\bf x})$, the Fourier transformed, perturbed Einstein 
equations are 
\begin{eqnarray}
\delta\ddot{\phi}_k &+& 3H\delta\dot{\phi}_k
+ \left[\frac{k^2}{a^2}+3\lambda 
(\phi^2+\langle \delta \phi^2 \rangle) 
+g^2(\chi^2+\langle \delta \chi^2 \rangle) \right] 
\delta\phi_k \nonumber \\
&=& 4\dot{\phi} \dot{\Phi}_k 
+ 2(\ddot{\phi}
+3H\dot{\phi})\Phi_k-2g^2\phi\chi \delta\chi_k,
\label{deltaphi}
\end{eqnarray}
\begin{eqnarray}
\delta\ddot{\chi}_k &+& 3H\delta\dot{\chi}_k+
\left[ \frac{k^2}{a^2}+g^2(\phi^2+\langle \delta \phi^2 \rangle)
\right] \delta\chi_k \nonumber \\
&=& 4\dot{\chi} \dot{\Phi}_k 
+ 2(\ddot{\chi}+3H\dot{\chi})\Phi_k-2g^2\phi\chi\delta\phi_k,
\label{deltachi}
\end{eqnarray}
\begin{eqnarray}
\dot{\Phi}_k+H\Phi_k=4\pi G
(\dot{\phi} \delta\phi_k+\dot{\chi} \delta\chi_k).
\label{Phi}
\end{eqnarray}
As long as $\delta\chi_k$ fluctuations in low momentum modes are not 
strongly  suppressed during inflation (i.e., $g^2/\lambda < 10$) 
and are excited during preheating, this 
leads to the growth of $\Phi_k$ and $\delta\phi_k$ on large scales, as is 
found in numerical simulations of Eqs.~(\ref{Phi}) and (\ref{deltaphi}).
Neglecting metric perturbations on the rhs of Eqs.~(\ref{deltachi}) which are
small during inflation, we find the following analytic solution
\begin{eqnarray}
\delta\chi_k=a^{-1} \left[c_1 \sqrt{\eta} H_{\nu}^{(2)} (k \eta) +c_2 
\sqrt{\eta} H_{\nu}^{(1)}(k \eta)\right],
\label{analyticchik}
\end{eqnarray}
with \cite{BV99}
\begin{eqnarray}
\nu^2=\frac94-\frac{g^2\phi^2}{H^2} \approx 
\frac94-\frac{3g^2}{2\pi \lambda}
\left(\frac{m_{\rm pl}}{\phi}\right)^2,
\label{nu2}
\end{eqnarray}
since $H^2 \approx 2\pi \lambda \phi^4/3$.
In the centre of the first resonance band, $g^2/\lambda=2$, $\nu^2$ is 
negative only when $\phi<2/\sqrt{3\pi} \sim 0.7m_{\rm pl}$, which means
that the exponential suppression can be avoided during most of 
inflation.  In this case large-scale metric perturbations are significantly 
amplified  during the preheating phase.

\begin{figure}
\epsfxsize = 3.5in
\epsffile{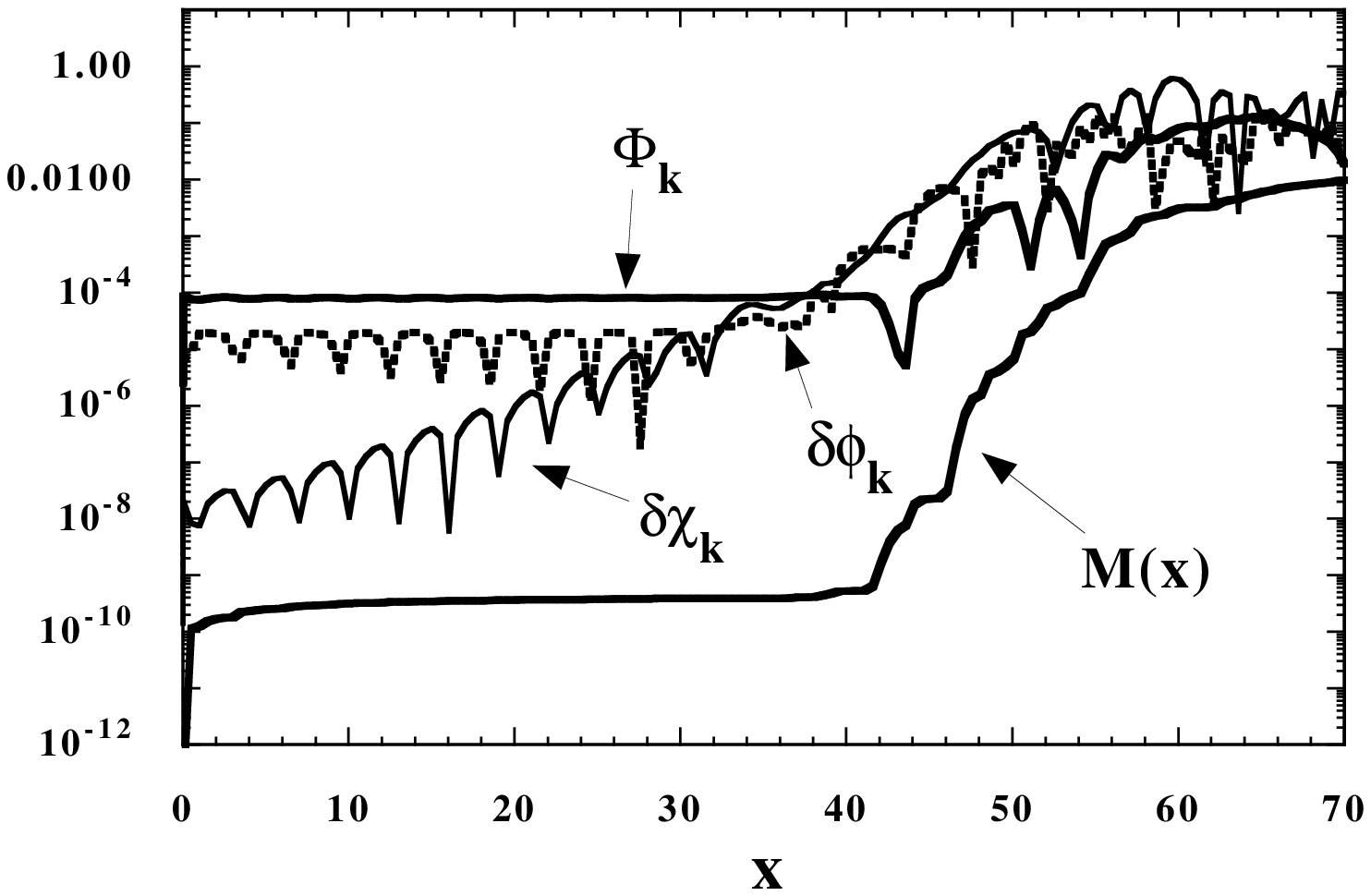}
\caption{The time evolution of 
$\Phi_k$, $\delta\chi_k$, $\delta\phi_k$, and
$M(x) \equiv \int_{x_i}^{x} \left( d\Phi/dx\right)^2 dx$ for
$g^2/\lambda=2$ during inflation and preheating for a cosmological mode.
In this case, the enhancement of metric perturbations leads to the 
production of magnetic fields due to the breaking of conformal flatness
of the background metric. } 
\label{metricevo}
\end{figure}

In Fig.~\ref{metricevo} we plot the evolution of $\Phi_k$, $\delta\chi_k$, 
$\delta\chi_k$, and $M(x) \equiv \int_{x_i}^{x} \left( d\Phi/dx\right)^2 
dx$ for $g^2/\lambda=2$ during inflation and preheating for a cosmological 
mode.  We include second order field and metric backreaction effects as 
spatial averages for background equations (see Refs.~\cite{ZBS,supp,PBH2} 
for details), and choose initial values for the scalar fields at the start of
inflation to be $\phi(0)=4m_{\rm pl}$ and 
$\chi(0)=10^{-3}m_{\rm pl}$.  Metric perturbations begin to grow during 
preheating after
$\delta\chi_k$ grows to or order $\delta\phi_k$, which results in the final 
amplitude of order $\Phi \sim 0.1$, clearly in conflict with  
observations of the CMB.

In spite of this, it is worth investigating this case 
in order to understand how the growth of metric perturbations  affects 
the evolution of magnetic fields.  The $M(x_f) =\int_{x_i}^{x_f} 
\left( d\Phi/dx\right)^2 dx$ term on the 
rhs of Eq.~(\ref{magestimation2}) becomes
of order $0.01$ (see Fig.~\ref{metricevo}), and the resulting 
magnetic field at 
decoupling is then estimated to be $|B_k^{\rm dec}|/ 1 {\rm G} 
\approx 10^{-4} 
a_i/a_{\rm dec}$.  When $a_i/a_{\rm dec}~\gsim~10^{-21}$ which corresponds 
to the reheating temperature, $T_R~\lsim~10^{11}$ GeV, magnetic fields 
exceed the value, $|B_k^{\rm dec}|~\sim~10^{-25}$ G, which is required to 
seed the galactic dynamo.

With the increase of $g^2/\lambda$, the inflationary suppression for long wave 
$\delta\chi_k$ modes begin to be significant.  For example, in the centre 
of the second resonance band, $g^2/\lambda=8$, the suppression is relevant 
for $\phi~\lsim~1.3m_{\rm pl}$.  In the Hartree approximation, the 
enhancement of super-Hubble metric perturbations during preheating was 
found to be weak for $g^2/\lambda~\gsim~8$ due to the suppressed $\chi$ 
fluctuation at the end of inflation \cite{ZBS}, which means that magnetic 
field fluctuations are hardly enhanced by large-scale metric perturbations.

However, since sub-Hubble $\delta\chi_k$ fluctuations are
free from strong inflationary suppression and exhibit parametric amplification
during preheating, metric preheating is typically vital on sub-Hubble 
scales \cite{PBH2}.  Then the mode-mode coupling between small-scale metric 
and large-scale magnetic field in Eq.~(\ref{perturbedmax}) may 
lead to the production of magnetic fluctuations.
In this case analytic estimations by Eq.~(\ref{magestimation}) can no longer 
be applied, and we have to solve the complicated nonlinear equation 
(\ref{perturbedmax}) directly.
Whether magnetic fields can be sufficiently amplified by the 
growth of small-scale metric perturbations is uncertain at present.  We 
leave to future work for the precise analysis of this issue.

We should also note that parametric excitation of sub-Hubble $\delta\chi_k$ 
modes will stimulate the growth of large-scale $\delta\phi_k$ and 
$\Phi_k$ modes.
The Hartree approximation misses this rescattering 
effect \cite{resca1,resca2}, which is expected to be important once 
fluctuations begin to be amplified significantly.  
In fact it was recently found that 
rescattering can lead to the amplification of super-Hubble metric 
perturbations even for $g^2/\lambda~\gsim~8$ in one-dimensional lattice 
simulations \cite{mpreFK}.  It is unknown whether this holds true for 
$g^2/\lambda \gg 1$, which will be clarified by fully nonlinear 
three-dimensional calculations.

It is certainly of interest to find parameter regions which satisfy both the 
CMB constraints and produce sufficient large-scale seed magnetic 
fields.  Although we have restricted ourselves in the chaotic inflationary 
scenario, the ratio $a_i/a_{\rm dec}$ and the energy scale of inflation are 
model-dependent. It is encouraging that we can test inflationary 
models by the  magnetic fields produced, together with CMB and 
primordial black hole  over-production constraints 
during preheating \cite{PBH1,PBH2}.

\section{Conclusions}

In this paper we have considered the amplification 
of (hyper-)magnetic fields during inflation and preheating. The conformal 
invariance of the standard Maxwell equations and the conformal flatness of
the FLRW background leave the observed cosmic magnetic fields as a major
mystery. 
In order to overcome such obstacles, we have considered three 
specific mechanisms:

(1) Couple the magnetic field to a coherently oscillating scalar field which
induces resonant growth of the magnetic field. 
In the presence of plasma effects,  parametric amplification of 
magnetic fields is typically counteracted by the growth of conductivity.
This competition is model dependent and the final outcome depends 
sensitively on the conductivity during inflation, the resonance
and thermalisation phases (see Figs 4,5).

(2) Break conformal invariance of Maxwell's equations through 
non-renormalisable  couplings to the curvature such as 
$R A_{\mu} A^{\mu}$.  When the corresponding coupling constant, $\beta$, 
is  negative, strong amplification of the magnetic field occurs during 
inflation.  As a result it is a promising mechanism, though some fine-tuning
may be required not to over-produce the magnetic fields by the end of 
preheating. For positive $\beta$ the produced field is too weak to 
be relevant even with the resonance from preheating. 

(3) Break the conformal flatness of the background metric. During metric 
preheating super-Hubble metric perturbations grow exponentially. The 
resulting growth of the Weyl tensor leads to amplification of the magnetic 
field, which while it is generic, is a complex, mode-mode, coupling problem.

It is certainly of interest to consider issues such as the 
non-equilibrium aspects 
of the problem and a detailed model of e.g., the GUT gauge group and couplings 
between the relevant gauge fields and the curvature/other fields,
which we leave to future work.

\section*{ACKNOWLEDGEMENTS}

The authors thank Peter Coles, Alexandre Dolgov,  
Fabio Finelli, Alan Guth, David Kaiser, Roy Maartens, Antonio Maroto, 
Ue-Li Pen, Jos\'e Senovilla, Dam Son, Alexei Starobinsky and Raul Vera 
for enlightening discussions and comments 
over the long  course of this project. 

BB, GP and FV thank the Newton 
Institute for support and hospitality during the program ``Structure 
Formation in the Universe".  BB thanks UCT, Cape Town for hospitality 
during early stages of this work.  FV acknowledges support from CONACYT 
scholarship Ref:115625/116522.  ST was supported by the Waseda University 
Grant for Special Research Projects.

\section*{Appendix: Magnetic fields with $R F_{\mu \nu}
F^{\mu \nu}$ interactions}

The 1-loop QED result in curved space includes terms of the form $R F_{\mu
\nu} F^{\mu \nu}$ together with similar terms involving $R_{\mu \nu}$ and
$R_{\mu \nu \alpha \beta}$. These are more complex to treat as resonance
systems because of periodic divergences. To illustrate this we consider
the Lagrangian
\beq
{\cal L}=\frac{R}{16\pi G} -
\frac{1}{4}\left(1+b\frac{R}{m_e^2}\right)
F_{\mu\nu}F^{\mu\nu}+{\cal L}_{{\rm inflaton}},
\label{F2term}
\eeq
where $b$ is a constant \footnote{The coefficient $b$ was calculated in 
\cite{DH} using perturbation theory in $R/m_e^2$.   
However, as pointed out in \cite{TW}, this result is not applicable in the 
early universe and $b$ is left as an arbitrary constant.}.

The equation of motion for the Fourier modes of $A_{\mu}$ are
\beq 
A_{ik}''+k^2A_{ik}+ \frac{b}{m^2_e + bR} 
R'A_{ik}'=0.
\label{F2eq}
 \eeq 
In the limit of $R \gg m_e^2$ (the one appropriate for the early universe 
\cite{TW}), the coefficient of $A_{ik}'$  becomes $R'/R$ and is 
independent of $b$.

Since $R$ oscillates through zero [see e.g., Eq.~(\ref{curvature2})], 
the equation is not amenable to simple numerical analysis. In this regard it
is similar to the evolution equation for the potential $\Phi$ in the
single, oscillating, scalar field case \cite{earlypre}. As discussed at the
end of  \cite{BV99}, the periodic singularities do not forbid resonance
bands. In the case of negative $b$ the possibility of efficient amplification 
during inflation exists due to the negative coupling instability.


\end{document}